\documentclass[twocolumn]{aastex631}

\usepackage{amsmath}

\newcommand{\F}{{\mathcal{F}}}
\newcommand{\asini}{a_{\mathrm{p}}}

\def\tinytt#1{\tiny{\textrm{#1}}}

\newcommand\Tstrut{\rule{0pt}{2.9ex}}       % "top" strut
\newcommand\Bstrut{\rule[-1.3ex]{0pt}{0pt}} % "bottom" strut
\newcommand\TBstrut{\Tstrut\Bstrut}  

\usepackage{soul}

\graphicspath{ {./Plots/} {./Figures/} {./figs/} {./} }

\begin{document}

\title{Search for continuous gravitational waves from unknown neutron stars in binary systems with long orbital periods in O3 data}

\author[0000-0002-1845-9309]{P. B. Covas}
\affiliation{Max Planck Institute for Gravitational Physics (Albert Einstein Institute) and Leibniz Universit\"at Hannover\\
D-30167 Hannover, Germany}
\email{pep.covas.vidal@aei.mpg.de}
\author[0000-0002-1007-5298]{M. A. Papa}
\affiliation{Max Planck Institute for Gravitational Physics (Albert Einstein Institute) and Leibniz Universit\"at Hannover\\
D-30167 Hannover, Germany}
\author[0000-0002-3789-6424]{R. Prix}
\affiliation{Max Planck Institute for Gravitational Physics (Albert Einstein Institute) and Leibniz Universit\"at Hannover\\
D-30167 Hannover, Germany}

%\input{git_tag.tex}
%\date{\commitDATE; \commitIDshort-\commitSTATUS}

\begin{abstract}
Gravitational waves emitted by asymmetric rotating neutron stars are
the primary targets of continuous gravitational-wave searches.
Neutron stars in binary systems are particularly interesting due to the
potential for non-axisymmetric deformations induced by a companion
star.
However, all-sky searches for unknown neutron stars in binary systems
are very computationally expensive and this limits their sensitivity and/or breadth.
In this paper we present results of a search for signals with
gravitational-wave frequencies between $50$ and $150$~Hz, from systems
with orbital periods between $100$ and $1\,000$ days and projected
semi-major axes between $40$ and $200$~light-seconds.
This parameter-space region has never been directly searched before.
We do not detect any signal, and our results exclude
gravitational-wave amplitudes above $1.25 \times 10^{-25}$ at
$144.32$~Hz with $95\%$ confidence.
Our improved search pipeline is more sensitive than any previous
all-sky binary search by about $75\%$.
\end{abstract}


%\maketitle
%\tableofcontents

\section{Introduction}
\label{sec:intro}

% Introduction and background
Continuous gravitational waves (CWs) from rotating neutron stars with a sustained quadrupole are one of the types of gravitational waves yet to be detected by ground-based detectors such as Advanced LIGO \citep[for a recent review see][]{KeithReview}. The required quadrupole in the neutron star can be sourced by different mechanisms \citep[see][]{jones2024multimessengerobservationsscienceenabled}, such as a mountain, characterized by the ellipticity of the neutron star (namely the difference between two moments of inertia not aligned with the rotational axis); r-modes, an oscillation mode of the star due to its rotation that can be unstable to gravitational-wave emission; Ekman-pumping \citep[][and references therein]{Singh:2016ilt}.

Continuous wave searches can be differentiated by the information we have about the emitting source: on one extreme there are targeted searches, where the source is a known pulsar; on the other extreme there are all-sky searches, where the source is completely unknown. Due to this lack of information, all-sky searches represent the most computationally demanding type of continuous wave search \citep{WETTE2023102880}. These searches are interesting since most of the neutron stars in our galaxy have not yet been discovered \citep{Pagliaro:2023bvi}.

Due to the large parameter space being investigated, all-sky searches usually employ semi-coherent methods, where the dataset is separated in smaller segments, which are analyzed coherently, and incoherently combined afterwards (see \citealt{Tenorio_2021} for a recent review). Although these methods are less sensitive than fully coherent ones, they significantly reduce the required computational budget \citep{PhysRevD.85.084010}. After an initial search with a certain coherence time, any remaining candidates can be followed-up with a longer coherence time in a number of subsequent stages \citep{Steltner:2023cfk,O3allskyLIGO,Dergachev:2022lnt} in order to increase their statistical significance. The main drawback of this procedure is that (assuming the same model throughout the procedure), due to the increased coherent time, robustness to unmodeled physical effects is decreased.

When the unknown neutron star is in a binary system the search problem becomes even more challenging, since additional parameters need to be accounted for \citep{Leaci:2015bka}. However, since more than half of the known millisecond pulsars are in binary systems \citep{Manchester_2005}, and, additionally, accretion offers an additional mechanism to generate the quadrupole needed for a detectable continuous wave, searching for signals from unknown neutron stars in binaries is of utmost importance.

\begin{figure*}
  \begin{center}
    \includegraphics[width=2.0\columnwidth]{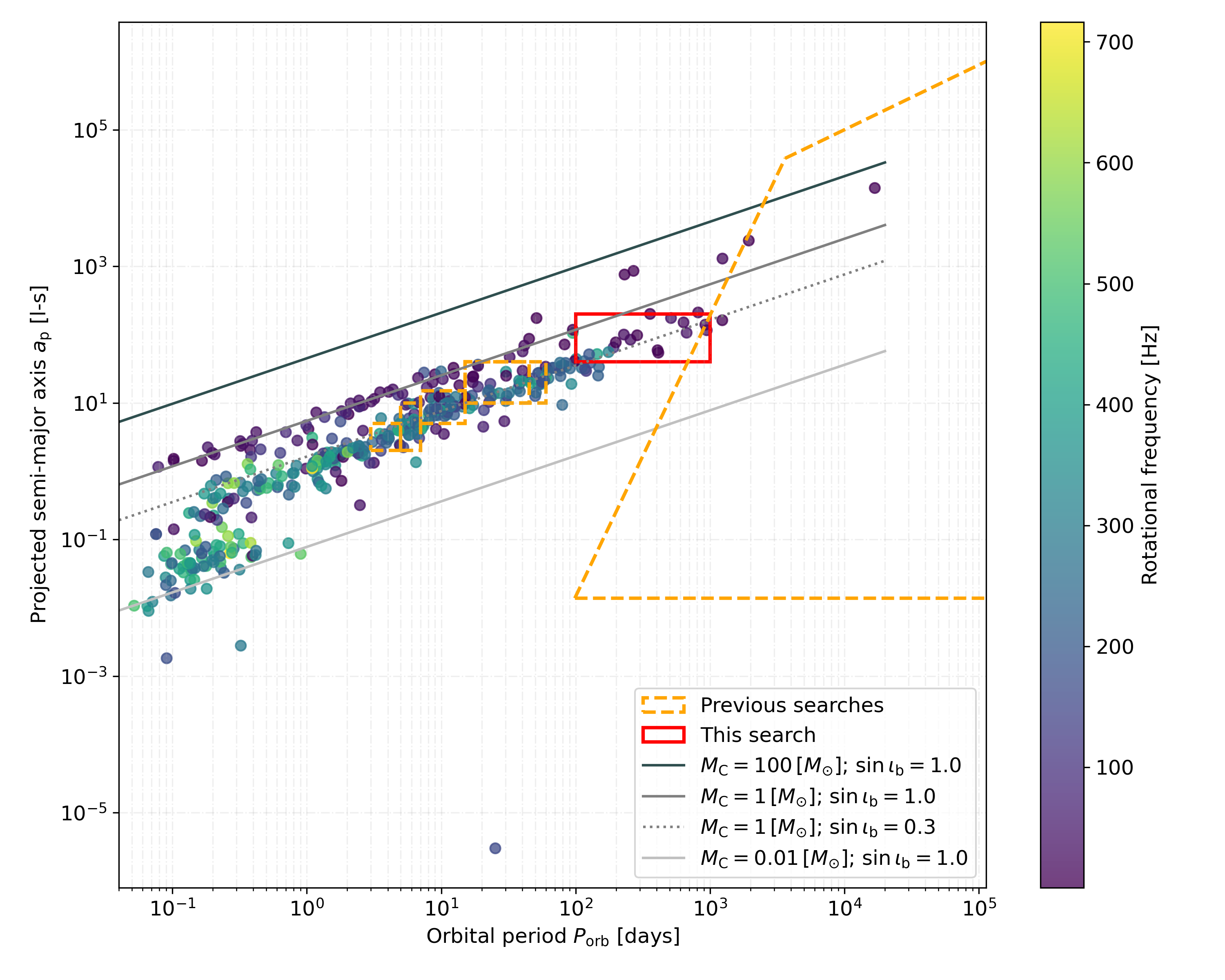}
    \caption{The red box shows the ranges in $( P_\mathrm{orb}, a_{\mathrm{p}} )$ covered by this search, while the points show the population of known pulsars in binary systems from the ATNF catalogue version 2.1.1 \citep{Manchester_2005}. The orange dashed boxes show the ranges covered by previous searches using Advanced LIGO data \citep{O2aallskybinary,O3aallskybinaryLIGO,Covas_2022}. The orange box to the right is the region where the results of \cite{Singh:2022hfd} apply, and partially overlaps with this search. The color scale shows the rotational frequency of each pulsar. The different gray lines show Kepler's third law for different values of the companion mass $M_\mathrm{C}$ and inclination of the binary orbit $\iota_\mathrm{b}$, assuming a neutron star mass of 1.4 M$_{\odot}$.}
    \label{fig:ParameterSpace}
  \end{center}
\end{figure*}

% Main new points: 1) region never searched before; 2) highest sensitivity ever reached for an all-sky binary search (similar to isolated all-sky searches)
In this paper we present a search with two main novelties: (i) the parameter space region that we investigate has never been directly searched before; (ii) the sensitivity depth of our search is $\sim 75\%$ greater than the most sensitive previous all-sky binary search \citep{O3aallskybinaryLIGO}. We cover gravitational-wave frequencies between 50 and 150 Hz, orbital periods between 100 and $1\,000$ days, and projected semi-major axes between 40 and 200 light-seconds (see Figure~\ref{fig:ParameterSpace}).

This search uses the public O3 data from the Advanced LIGO detectors, and attains a sensitivity depth of $38.6$ $\textrm{Hz}^{-1/2}$. We use the most sensitive search pipeline available for this type of search, \textsc{BinarySkyHou$\mathcal{F}$} \citep{BSHFstat}, which combines the coherent $\mathcal{F}$-statistic \citep{jaranowski_data_1998} with the Hough transform method \citep{krishnan04:_hough}. Although we do not detect any signal, we estimate upper limits on the gravitational-wave amplitude at a level of $1.25 \times 10^{-25}$ at $144.32$ Hz with $95\%$ confidence. Furthermore, due to the low coherence time of the main search and of the follow-ups ($0.05$ days to $1.56$ days), this search is robust to unmodeled effects such as spin-wandering \citep{spinwandering}, neutron star glitches \citep{PhysRevD.96.063004}, or timing noise \citep{PhysRevD.91.062009}, which may become important on the timescale of many days (see Section \ref{sec:robustness} for an extended discussion).

% Outline of the paper
The paper is organized as follows: in Section~\ref{sec:signal} we introduce the signal model that we assume; in Section~\ref{sec:method} we describe the data that we use, the parameter space that we cover, the pipeline that we use, and the search that was done; in Section~\ref{sec:results} we present our results and discuss their astrophysical implications; in Section~\ref{sec:conclusions} we conclude the paper.

\section{Signal model}
\label{sec:signal}

We assume that the continuous wave signal $h(t)$ as a function of time $t$ in the detector frame is given by \citep{jaranowski_data_1998}:
\begin{eqnarray}
  \label{eq:1}
  h(t) = h_0 & \left[ F_+(t; \psi) \frac{1 + \cos^2{\iota}}{2} \cos{[\phi_0 + \phi(t)]} \right. \nonumber \\
  + & \left. F_{\times}(t; \psi)  \cos{\iota} \sin{[\phi_0 + \phi(t)]} \right],
\end{eqnarray}
where $F_+$ and $F_{\times}$ are the antenna-pattern functions of the detector \citep[given in][]{jaranowski_data_1998}, $\iota$ is in the angle between the angular momentum of the neutron star and the line of sight, $\psi$ is the polarization angle, $\phi_0$ is the initial phase, $\phi(t)$ is the gravitational-wave phase at time $t$, and $h_0$ is the intrinsic gravitational-wave amplitude.

The phase of the signal $\phi(t)$ in the detector frame depends on the intrinsic gravitational-wave frequency $f_0$ and frequency derivatives of the signal $f_1,...,f_k$, and on the Doppler modulation due to the relative motion between the neutron star and the detector. This Doppler modulation depends on the sky position of the source (given by $\alpha$ and $\delta$) and on the orbit of the neutron star around the binary barycenter, described by the orbital period $P_\mathrm{orb}$, the projected semi-major axis $a_{\mathrm{p}}$, the time of ascension $t_{\mathrm{asc}}$, the argument of periapsis $\omega$, and the eccentricity $e$, as discussed by \cite{Leaci:2015bka}. Here we search for sources with a small enough spin-down and eccentricity to not require that we search over these parameters (i.e. a single value is used) in order to accurately track the signal in the first stage of the search, as reflected in Table~\ref{tab:region}.

\section{The search}
\label{sec:method}

\subsection{Data}
\label{sec:data}

This search uses O3 public data \citep{GWOSCO3a,GWOSCO3b} of the Advanced LIGO Hanford (H1) and Advanced LIGO Livingston (L1) gravitational-wave detectors \citep{LIGOScientific:2014pky}. The data-span covers from April 2019 to March 2020, with a month-long interruption for commissioning interventions. The average duty factor is $\sim 74\%$. We use the \texttt{GWOSC-16KHZ\_R1\_STRAIN} channel and the \texttt{DCS-CALIB\_STRAIN\_CLEAN\_SUB60HZ\_C01\_AR} frame type.

A linear and non-linear noise subtraction procedure was applied to this data before public release at the harmonics of the mains 60\,Hz lines and the calibration lines \citep{Davis:2018yrz} in order to remove some of the non-Gaussian sinusoidal-like noise features present in the dataset \citep{PhysRevD.97.082002}. Many other lines still contaminate the dataset, which produce spurious outliers that need to be individually examined and discarded, as will be seen in Section~\ref{sec:cands}.

% Gating
This dataset also suffers from a large number of short-duration glitches that increase the noise level in the frequency range of interest for this search, thus reducing its sensitivity, as discussed in other publications \citep[e.g.][]{O3allskyLIGO}. For this reason, we use the gating method developed by \citet{gating} to remove these glitches and improve data quality.

\begin{figure}
  \begin{center}
    \includegraphics[width=1.0\columnwidth]{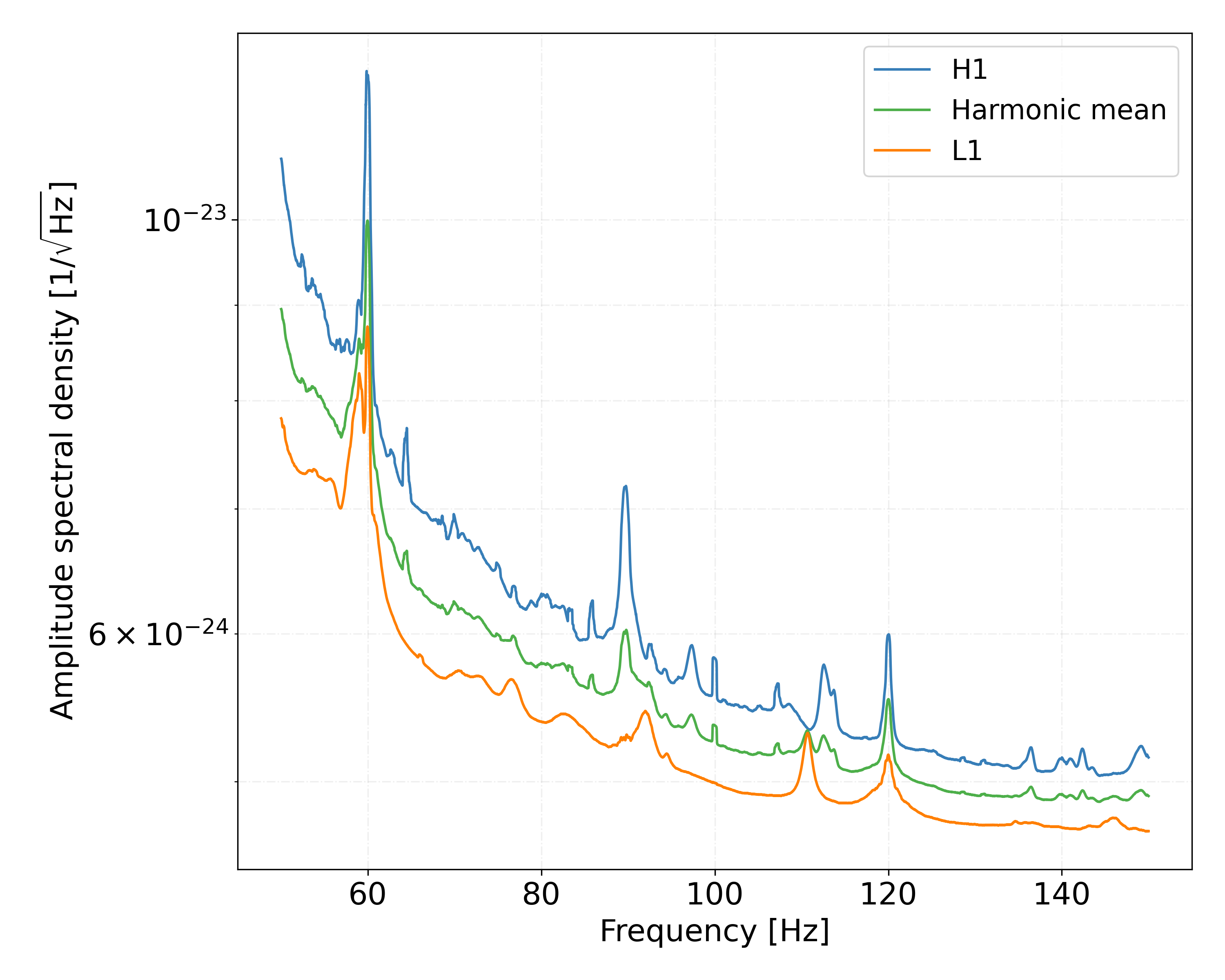}
    \caption{Harmonic mean (over $105\,111$ and $105\,269$ SFTs of 200 s for H1 and L1, respectively) of the amplitude spectral density $\sqrt{S_n}$ of the data used in this search as a function of frequency. The upper blue curve shows H1, the lower orange curve shows L1, and the middle green curve shows the harmonic mean between the two.}
    \label{fig:PSD}
  \end{center}
\end{figure}

The dataset of each detector is divided in short Fourier transforms \citep[SFTs, see][]{Allen_Mendell} of 200 s, short enough so that the signal power does not spread by more than a frequency bin during this time. The total number of SFTs is $105\,111$ for H1 and $105\,269$ for L1. The harmonic mean amplitude spectral density $\sqrt{S_n}$ of the Advanced LIGO detectors in the frequency range covered by this search is shown in Figure~\ref{fig:PSD}.

\subsection{Parameter space covered}
\label{sec:space}

The signal parameter space covered by the search is summarized in Table~\ref{tab:region}. We search for signals across the entire sky with gravitational-wave frequencies $f_0$ between 50 and 150\,Hz and frequency derivatives $|f_1| \le 7.6 \times 10^{-12}$\,Hz/s. The frequency range is chosen as a balance between the sensitivity of the detectors (see Figure~\ref{fig:PSD}), the computational cost of the search (which grows $\propto f_0^5$), and the observed rotational frequencies of the pulsar population (5 out of 17 known pulsars in our $( P_\mathrm{orb}, a_{\mathrm{p}} )$ range fall in this frequency range), as shown in Figure~\ref{fig:ParameterSpace}. The chosen frequency derivative range allows us to set $f_1 = 0$ Hz/s for all templates in the first stage of the search while keeping the average mismatch below $\lesssim 0.3$. With the most rapidly varying rotation frequency of any millisecond pulsar in a binary system being $\sim 10^{-14}$ Hz/s \citep{Manchester_2005}, our search range comfortably covers the observed range.

Neutron stars in binary systems display a broad range of projected semi-major axes $\asini$, depending on the mass of the companion $M_\mathrm{C}$, on the inclination angle of the orbital plane $\iota_\mathrm{b}$, and on the orbital period $P_\mathrm{orb}$, as shown in Figure~\ref{fig:ParameterSpace}. All-sky searches with high sensitivity over such a broad range of parameters are currently impossible due to their prohibitive computational cost, so we limit the search to the range in $( P_\mathrm{orb}, a_{\mathrm{p}} )$ indicated by the red box in that Figure and given in Table~\ref{tab:region}. The parameter space contained in this box has not been directly searched in any previous search \citep[][]{TwoSpectS6,O2aallskybinary,Covas_2022,O3aallskybinaryLIGO, Erratum}. The neutron stars targeted by this search have companions with masses on the order of a solar mass (mostly helium white dwarfs), and typically present smaller rotational frequencies than those inhabiting other regions of the $( P_\mathrm{orb}, a_{\mathrm{p}} )$ space. It has been shown in \citet{Wang:2022duk} that neutron stars with millisecond rotation rates can be produced in systems with orbital periods between 50 and 1200 days.

Finally, in the first stage of the search we also set all templates to $e = 0$ (which still covers a range in eccentricity given by equation \eqref{eq:eccmax}) and do not explicitly search over eccentricity (and argument of periapsis).

Assuming small eccentricity, we estimate the maximum eccentricity covered by
  this search using Equations~(15), (62), and (64) of
  \citet{Leaci:2015bka} with a {\it{metric}} mismatch of $\mu=1$:
\begin{equation}
  e_{\mathrm{max}} = 0.08\,\left({150\,\textrm{Hz}\over{f_0}}\right) \left({{P_\mathrm{orb}} \over 1\,000\,\textrm{days}}\right) \left({200\,\textrm{l-s}\over{a_{\mathrm{p}}}}\right).
  \label{eq:eccmax}
\end{equation}
For those ($f_0,P_\mathrm{orb},a_{\mathrm{p}}$) that yield large $e_{\mathrm{max}}$ values, however, the assumptions of Equation~\ref{eq:eccmax} do not hold and can give rise to {\it{observed}} mismatches larger than expected. As a pragmatic fix, we cap Equation~\ref{eq:eccmax} at $e_{\mathrm{max}} \leq 0.1$.

\begin{deluxetable}{lc}
\label{tab:region}
\tablecaption{Range of values for the different parameters covered by the search. Only a single template is required by the first search to cover the ranges in first frequency derivative $f_1$, in eccentricity $e$, and in the argument of periapsis $\omega$. $t_m$ is the mid-time of the search, equal to $1\,253\,764\,659.5$ (GPS), which is used as the reference time.}
\tablehead{\colhead{Parameter} & \colhead{Range}}
\startdata
\TBstrut $f_0$ [Hz] & [50, 150] \\ 
\TBstrut $|{f_1}|$ [Hz/s] & $ < 7.6 \times 10^{-12}$ \\ 
\TBstrut $\alpha$ [rad] & [0, $2\pi$) \\ 
\TBstrut $\delta$ [rad] & [$-\pi/2$, $\pi/2$] \\
\TBstrut $a_{\mathrm{p}}$ [l-s] & [40, 200] \\ 
\TBstrut $P_\mathrm{orb}$ [days] & [100, $1\,000$] \\ 
\TBstrut $t_{\mathrm{asc}}$ [GPS] & [$t_m - P_\mathrm{orb}/2$, $t_m + P_\mathrm{orb}/2$) \\ 
\TBstrut $e$ & $ < \min\left[0.1, e_{\max{}}\right] $ (see Equation~\eqref{eq:eccmax}) \\
\TBstrut $\omega$ [rad] & [0, $2\pi$) \\  
\enddata
\end{deluxetable}

\subsection{Target signal population}
\label{sec:fakeSignals}
Various parameters of the search, such as follow-up thresholds and veto parameters, are chosen to ensure safety (i.e. very low false dismissal probability) of the target signal population. The parameters of this target population are drawn from the search ranges given in Table~\ref{tab:region}, with uniform distributions on $\cos\iota$, $\psi$ and $\phi_0$ from Equation~\eqref{eq:1}.
All search parameters are sampled uniformly, except for sky position, which is drawn isotropically. The frequency $f_0$ is drawn uniformly in ten different 0.1 Hz bands ($56.3,59.3,63.2,77.5,83.1,85.5,100.1,101.5,131.3,148.2$). Spin-down $f_1$ and eccentricity $e$ are log-uniformly distributed with a lower limit two orders of magnitude below the maximum given in Table~\ref{tab:region}.

The amplitudes $h_0$ are chosen by uniformly sampling the sensitivity depth (see Equation~\eqref{eq:sensDeph}) from the range $[15, 110]$~Hz$^{-1/2}$ (in discrete steps of $5$~Hz$^{-1/2}$). We use a total set of 50\,000 signals, with 250 signals for each frequency and sensitivity depth. For the follow-up tests presented in Section~\ref{sec:fu}, we use a subset of 2\,000 signals (10 signals for each frequency and sensitivity depth) due to the high computational cost of these tests.

\subsection{Stage 0}
\label{sec:pipeline}

For the first stage of the search, which scans the entire parameter space, we use a semi-coherent search method where the data is separated in segments of a span $T_{\mathrm{coh}} = 4\,500$\,s, a value selected to balance computational cost constraints and sensitivity. We use \textsc{BinarySkyHou$\mathcal{F}$} \citep{BSHFstat}, which currently is the most sensitive pipeline for these types of searches.

For each segment $\ell=1,\dots, N_{\mathrm{seg}}$ we calculate the coherent multi-detector dominant-response detection statistic $\mathcal{F}_{\mathrm{AB};\ell}$ \citep{2DOF} over a coarse grid in $f$ and sky. Using $\mathcal{F}_{\mathrm{AB}}$ instead of the classic $\mathcal{F}$-statistic yields sensitivity gains for these short segments due to the reduced number of degrees of freedom.

This is the longest coherence time $T_{\mathrm{coh}}$ used so far in an all-sky search for neutron stars in binary systems, resulting in the most sensitive search.
However, it is short enough so that the orbital parameters $( P_{\mathrm{orb}}, a_{\mathrm{p}}, t_{\mathrm{asc}} )$ are not resolved at the coherent stage.

By combining the information from the different $\mathcal{F}_{\mathrm{AB};\ell}$ over the entire observation time, we can refine our ability to localize where a signal comes from in the sky, as well as the values of its orbital parameters. In practice this means that the coherent grid used for the single-segment calculations is not fine enough for combining together those results.

We set up a finer sky-grid and grids in $\lambda^{\mathrm{orb}}\equiv( P_{\mathrm{orb}}, a_{\mathrm{p}}, t_{\mathrm{asc}} )$, and compute the detection statistic at each point by adding together appropriately the results computed in the coherent coarse-grid template. The Hough master equation specifies what ``appropriately'' means as it determines at every segment which is the closest coarse template $\lambda^{\mathrm{coarse}}_\ell=(f^{\textrm{coarse}}_\ell,{\textrm{sky}}^{\textrm{coarse}}_\ell)$ to the template ($f_0, \lambda^{\mathrm{orb}}, \textrm{sky}$) for which we wish to compute the detection statistic \citep{BSHFstat}:
\begin{equation}
\label{eq:twoFsum}
  \hat{\mathcal{F}}_{\mathrm{ABw}}(f_0, \lambda^{\mathrm{orb}} , \textrm{sky})  \equiv \sum_{\ell=1}^{N_{\mathrm{seg}}} w_\ell \,\mathcal{F}_{\mathrm{AB};\ell}(\lambda^{\mathrm{coarse}}_\ell),
\end{equation}
where $N_{\mathrm{seg}}=5\,910$ and
\begin{equation}
  w_\ell \propto \begin{cases} \frac{A_\ell + \frac{C_\ell^2}{A_\ell}}{S_{\mathrm{n;}\ell}} & \text{if } A_\ell \ge B_\ell, \\
                            \frac{B_\ell + \frac{C_\ell^2}{B_\ell}}{S_{\mathrm{n;}\ell}} & \text{otherwise}.
  \end{cases}
\end{equation}
$A$, $B$, and $C$ are the antenna-pattern matrix coefficients \citep{jaranowski_data_1998} calculated at each coarse sky point. The weights are normalized as $\sum_\ell w_\ell = N_{\mathrm{seg}}$.

For each coarse sky-grid point we do not compute $\mathcal{F}_{\mathrm{AB};\ell}$ for segments with $w_\ell$ values in the 25th percentile and lower \citep{BSHFstat,Behnke:2014tma}. This reduces the computational cost of the search while incurring a loss of sensitivity smaller than $\sim 5\%$. All segments are however used to recalculate the detection statistic for the top $1\%$ templates of each frequency band.

We define a significance $s$ (also called critical ratio) for every template as
\begin{equation}
  s = \frac{ \hat{\mathcal{F}}_{\mathrm{ABw}} - \mu_\F}{\sigma_\F},
\label{eq:sig}
\end{equation}
where $\mu_\F$ is the expected mean and $\sigma_\F$ is the standard deviation of $\hat{\mathcal{F}}_{\mathrm{ABw}}$ in Gaussian noise, given by:
\begin{equation}
  \mu_\F = 2 N_{\mathrm{seg}}, \quad \sigma_\F^2 = 4 \sum\limits_{\ell=1}^{N_{\mathrm{seg}}} w^2_\ell.
  \label{eq:sigma}
\end{equation}

\begin{figure}[htbp]
    \centering
    \includegraphics[width=\columnwidth]{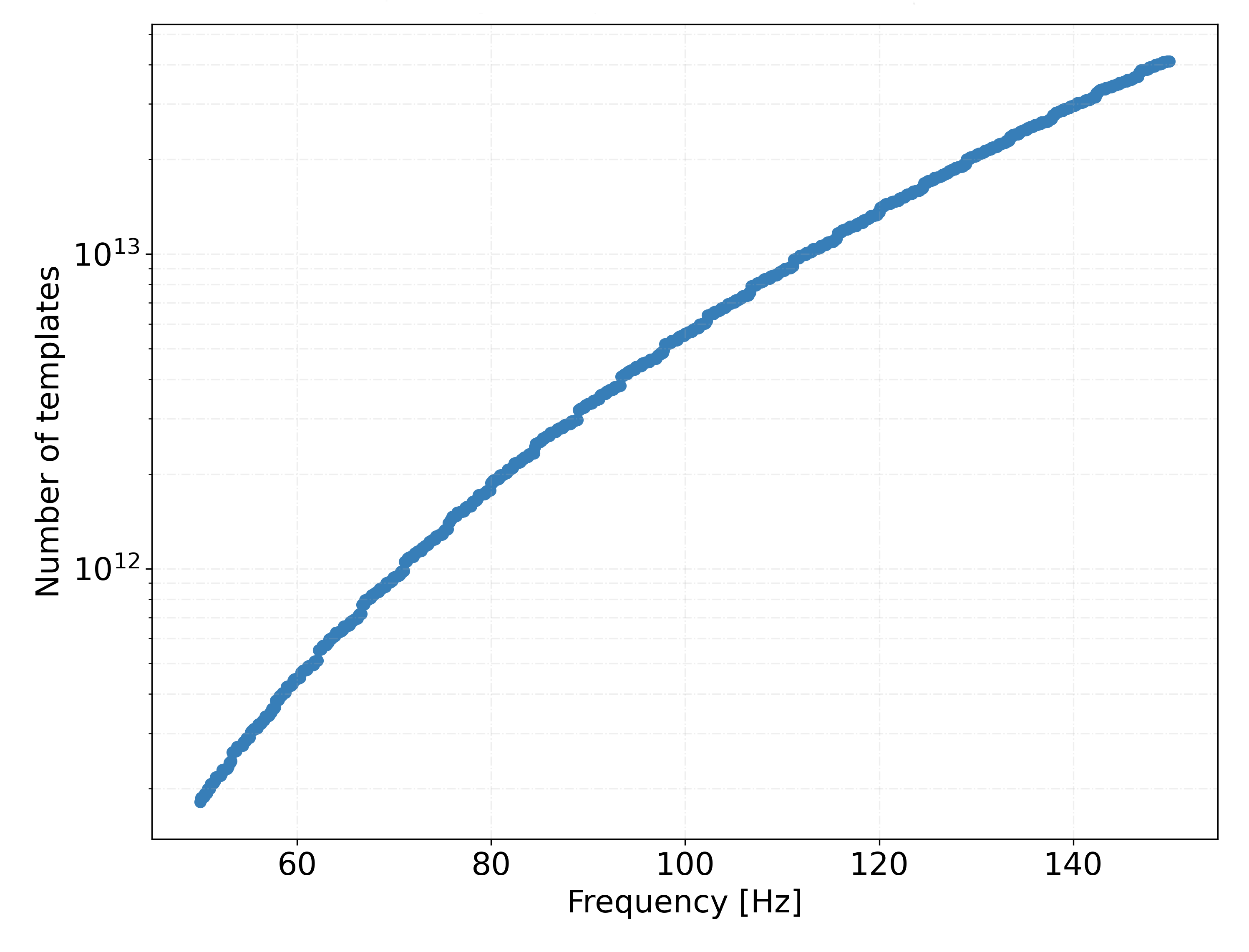}
    \caption{Number of templates searched in each 0.1 Hz frequency band as a function of frequency. The total number of templates is $\sim 1.05 \times 10^{16}$.}
    \label{fig:ntemplates}
\end{figure}

\begin{figure}
  \begin{center}
    \includegraphics[width=1.0\columnwidth]{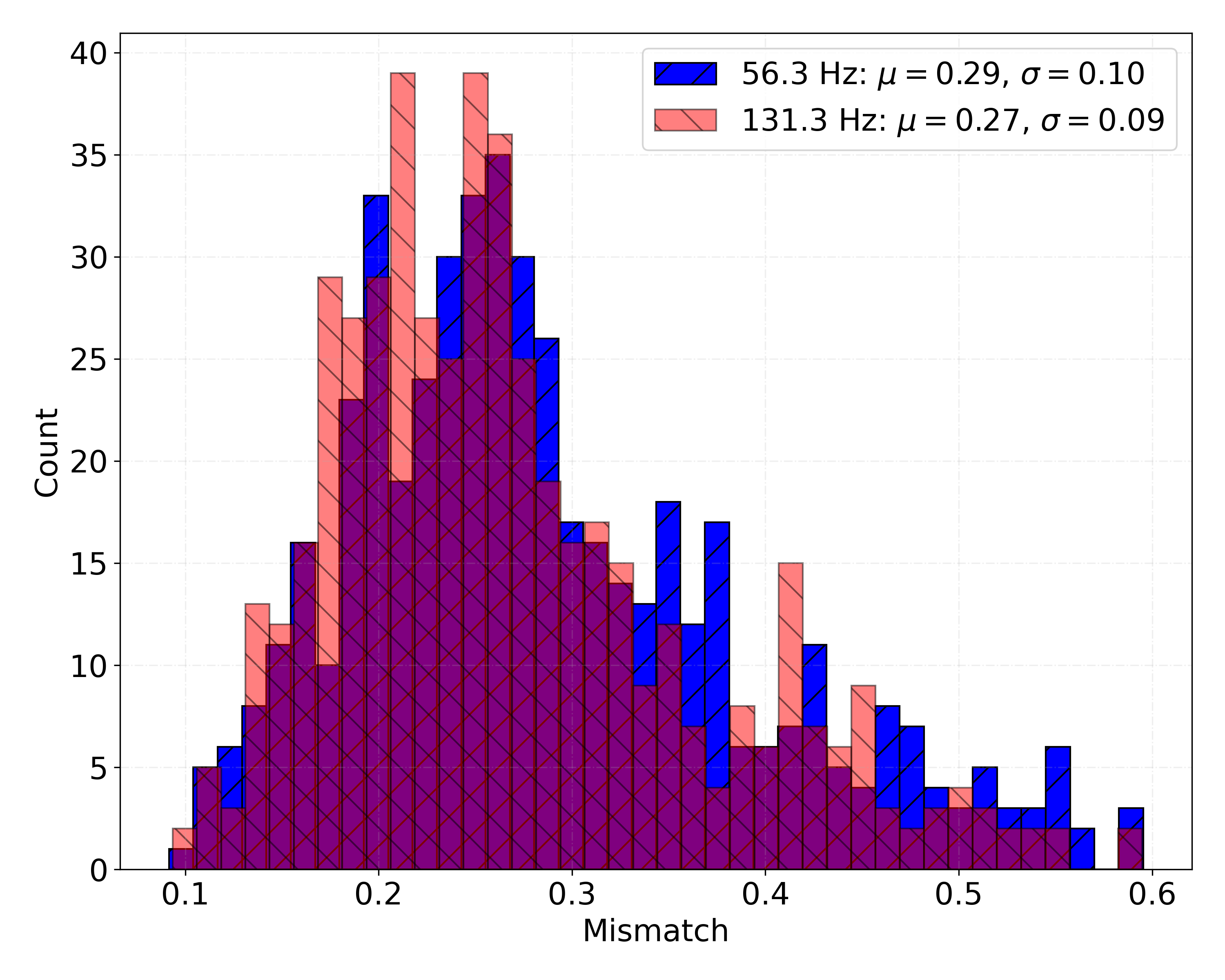}
    \caption{Mismatch distribution of this search at two different 0.1 Hz frequency bands, with the other parameters that describe the continuous wave signal spanning the ranges given in Table~\ref{tab:region}. The legend shows the mean and standard deviation of each distribution.}
    \label{fig:Mismatch}
  \end{center}
\end{figure}

The search is divided in frequency bands of 0.1 Hz. The total number of templates searched is $\sim 1.05 \times 10^{16}$, which are shown for each of the 0.1 Hz bands in Figure~\ref{fig:ntemplates}. The grid resolutions for all the parameters are shown in Table~\ref{tab:setup}. These resolutions are obtained by using Equations 30-32 of \citet{2019PhRvD..99l4019C} with metric mismatch $\mu=1$ in each dimension; Equation 34 with $P_\mathrm{F} = 0.8$; and half of the value given in Equation 33 for $f_0$. The overall resulting average mismatch produced by our template grid is $\lesssim 0.3$ and the mismatch distributions at two different frequencies are shown in Figure~\ref{fig:Mismatch}. We use finer coherent and semi-coherent grids for this search in order to increase its sensitivity compared to previous searches of this kind such as \citet{Covas_2022}.

\begin{deluxetable}{lc}
\label{tab:setup}
\tablecaption{Grid resolutions for the different parameters in the first stage of the search. $\Omega=2\pi/P_\mathrm{orb}$ is the average angular orbital velocity.}
\tablehead{Parameter & Resolution}
\startdata
\TBstrut $\delta f_0    $ [Hz]  & $ 1.1 \times 10^{-4} $ \\
\TBstrut $\delta \alpha $ [rad] & $ 1.9 \times 10^{-2} \, \left({{150 \, \tinytt{Hz}}\over{f_0}}\right)$ \\
\TBstrut $\delta \delta $ [rad] & $ 1.9 \times 10^{-2} \, \left({{150 \, \tinytt{Hz}}\over{f_0}}\right)$ \\
\TBstrut $\delta \asini $ [l-s]  & $ 3.2 \times 10^{1} \, \left({{150 \, \tinytt{Hz}}\over{f_0}}\right) \left({{P_\mathrm{orb} \over {1\,000 \, \tinytt{days}}}}\right) $ \\
\TBstrut $\delta \Omega $ [Hz]  & $ 1.8 \times 10^{-8} \, \left({{150 \, \tinytt{Hz}}\over{f_0}}\right) \left({{P_\mathrm{orb} \over {1\,000 \, \tinytt{days}}}}\right) \left({{200 \, \tinytt{l-s}}\over{\asini}}\right) $ \\
\TBstrut $\delta t_{\mathrm{asc}} $ [GPS] & $ 2.2 \times 10^{6} \, \left({{150 \, \tinytt{Hz}}\over{f_0}}\right) \left({{P_\mathrm{orb} \over {1\,000 \, \tinytt{days}}}}\right)^2 \left({{200 \, \tinytt{l-s}}\over{\asini}}\right)$ \\
\enddata
\end{deluxetable}

\subsection{Candidate selection}
\label{sec:cands}

\begin{figure}
  \begin{center}
    \includegraphics[width=1.0\columnwidth]{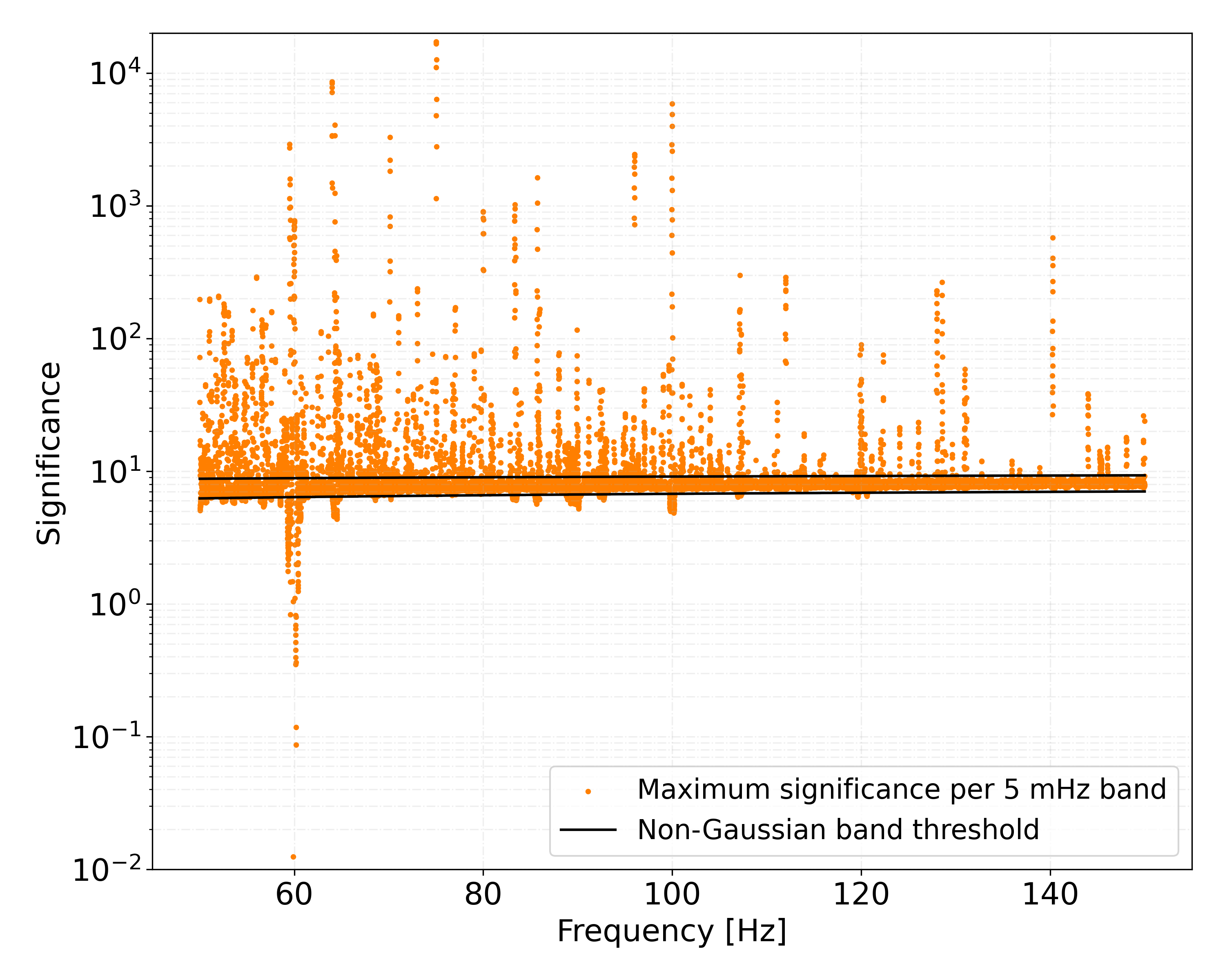}
    \caption{Maximum significance $s$ in each 5 mHz frequency sub-band as a function of frequency. Significance values well below the average level are due to spectral disturbances that lead to an ``under-weighing" of the data from various normalization procedures and produce low values of the detection statistic. The black lines are used to determine the frequency bands where the upper limits apply, see Section~\ref{sec:UL}.}
    \label{fig:Toplists}
  \end{center}
\end{figure}

We divide each 0.1 Hz band in 5 mHz sub-bands and consider the $10^6$ templates with the highest significance $s$ in every sub-band. Compared to previous searches we have not picked the top candidates directly from the 0.1 Hz bands because the low-frequency region is affected by very loud but narrow spectral disturbances, whose contamination can be effectively contained by simply choosing the top candidates in smaller bands. Figure~\ref{fig:Toplists} shows the significance $s$ of the top template in each 5 mHz sub-band.

We use the clustering procedure of \citet{2019PhRvD..99l4019C,O2aallskybinary} to group together results attributed to the same origin. The cluster with the highest significance per 5 mHz band is saved for further analysis, yielding a total of 20\,000 clusters. In what follows we refer to these selected clusters as {\it{candidates}}.

\begin{figure}
  \begin{center}
    \includegraphics[width=1.0\columnwidth]{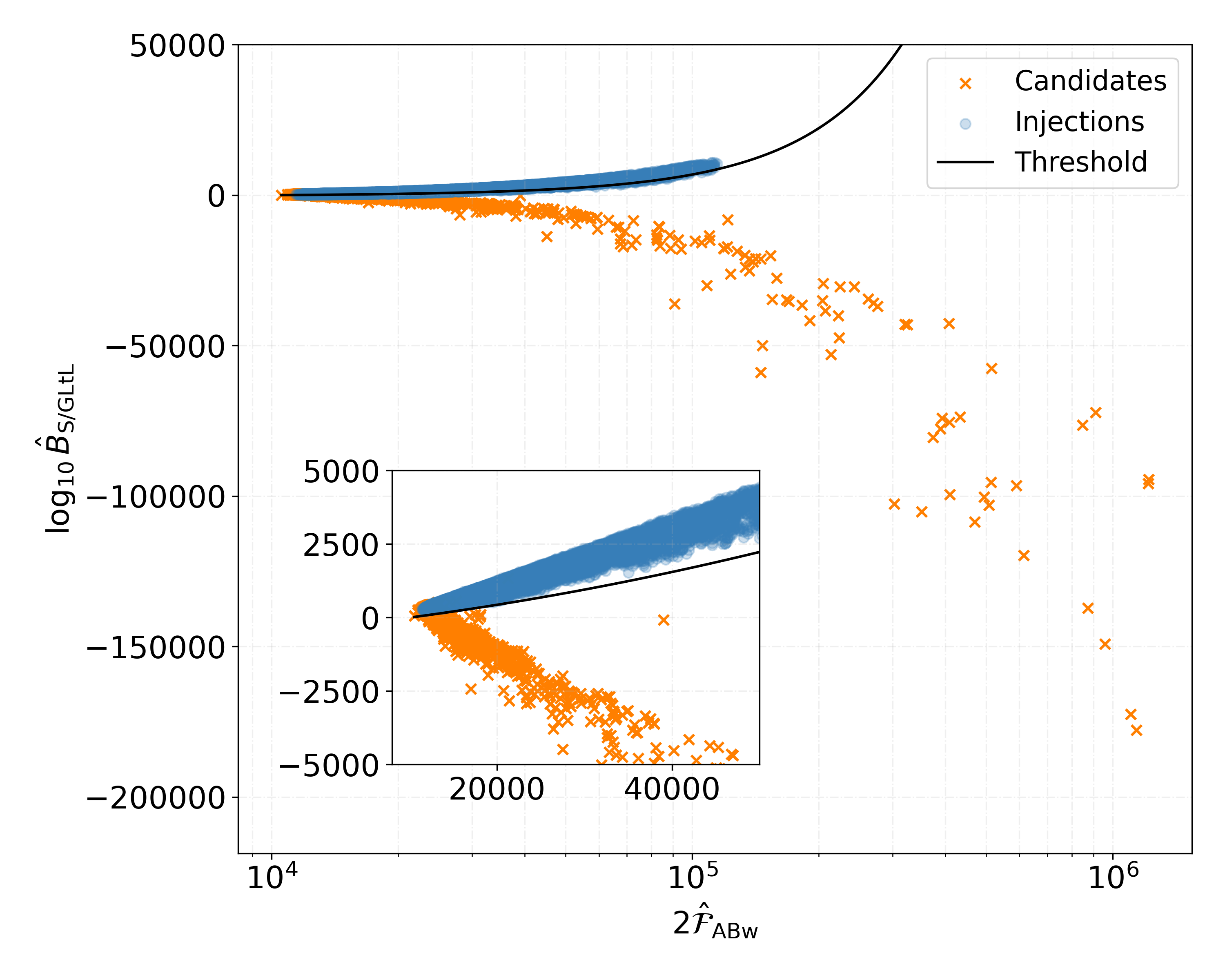}
    \caption{Distribution of $\log_{10} \hat{B}_\mathrm{S/GLtL}$ values as a function of $2\hat{\mathcal{F}}_{\mathrm{ABw}}$ for 25\,673 injected signals from our target population (shown with blue circles) and 20\,000 candidates (shown with orange crosses). The black line shows the threshold, given by $\log_{10} \hat{B}_\mathrm{S/GLtL} = x_0 + x_1 2\hat{\mathcal{F}}_{\mathrm{ABw}} + x_2 2\hat{\mathcal{F}}_{\mathrm{ABw}}^2$, with $x_0=-360.3$, $x_1= 3.1 \times 10^{-2}$, and $x_2 = 4.1 \times 10^{-7}$. The upper region is the candidate acceptance region.}
    \label{fig:BSGLtL}
  \end{center}
\end{figure}

Before analyzing the candidates with a longer coherence time, we check whether they might be due to a line. We do this by computing the $\log_{10} \hat{B}_\mathrm{S/GLtL}$ statistic of \cite{PhysRevD.93.084024} with $\hat{\mathcal{F}}_* = 0$. This effectively results in a comparison between the multi-detector and the largest of the single-detector $\mathcal{F}$-statistic values. For astrophysical signals we expect the multi-detector statistic to be larger than the single-detector. We calibrate the veto based on injected signals from our target population added to the real data (we only use signals that generate a cluster with a higher significance than the top cluster in their respective 5 mHz band), as illustrated in Figure~\ref{fig:BSGLtL}. This veto has a false dismissal probability of less than 0.004\%. Applied to our 20\,000 candidates, it removes 1\,718, leaving 18\,282 candidates for the next stage (most of the removed candidates are in the region with low $2\hat{\mathcal{F}}_{\mathrm{ABw}}$ values). $81.8\%$ of these removed candidates have a frequency smaller than 100 Hz, since that is the frequency region where most of the non-Gaussian disturbances are located.

\subsection{Follow-up}
\label{sec:fu}

In the next step we apply a follow-up procedure where the coherence time $T_{\textrm{coh}}$ is increased, and use a nested sampling algorithm \citep{bilby} to calculate the detection statistic \citep{covas2024new}. Due to the increased coherence time, the follow-up resolves different values of the frequency derivative, eccentricity, and argument of periapsis, which are now searched for explicitly in the ranges given by Table~\ref{tab:region}. We use the injected signals from our target population to determine the search region \citep{covas2024new}.

\begin{figure*}
  \begin{center}
    \includegraphics[width=1\columnwidth]{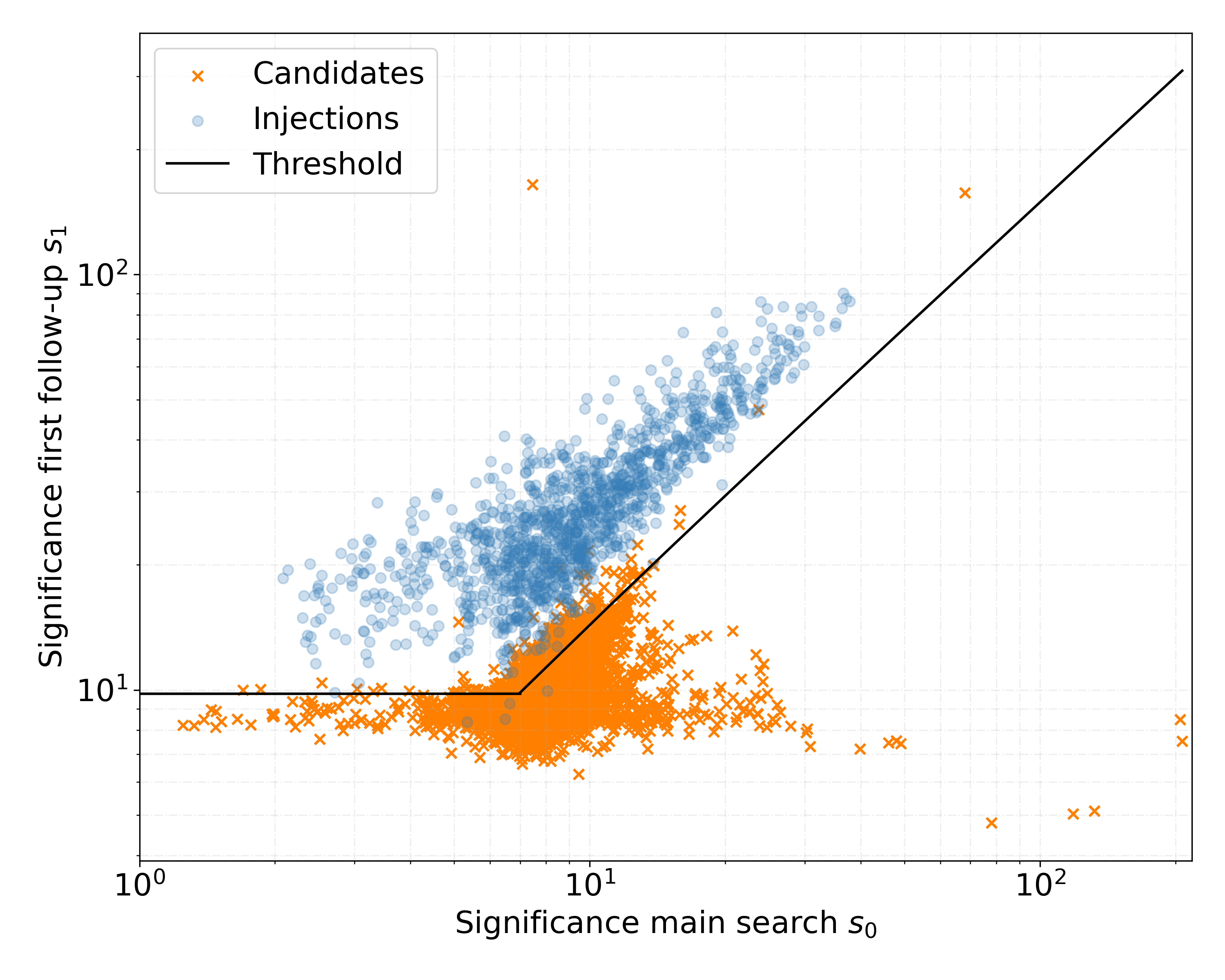}
    \includegraphics[width=1\columnwidth]{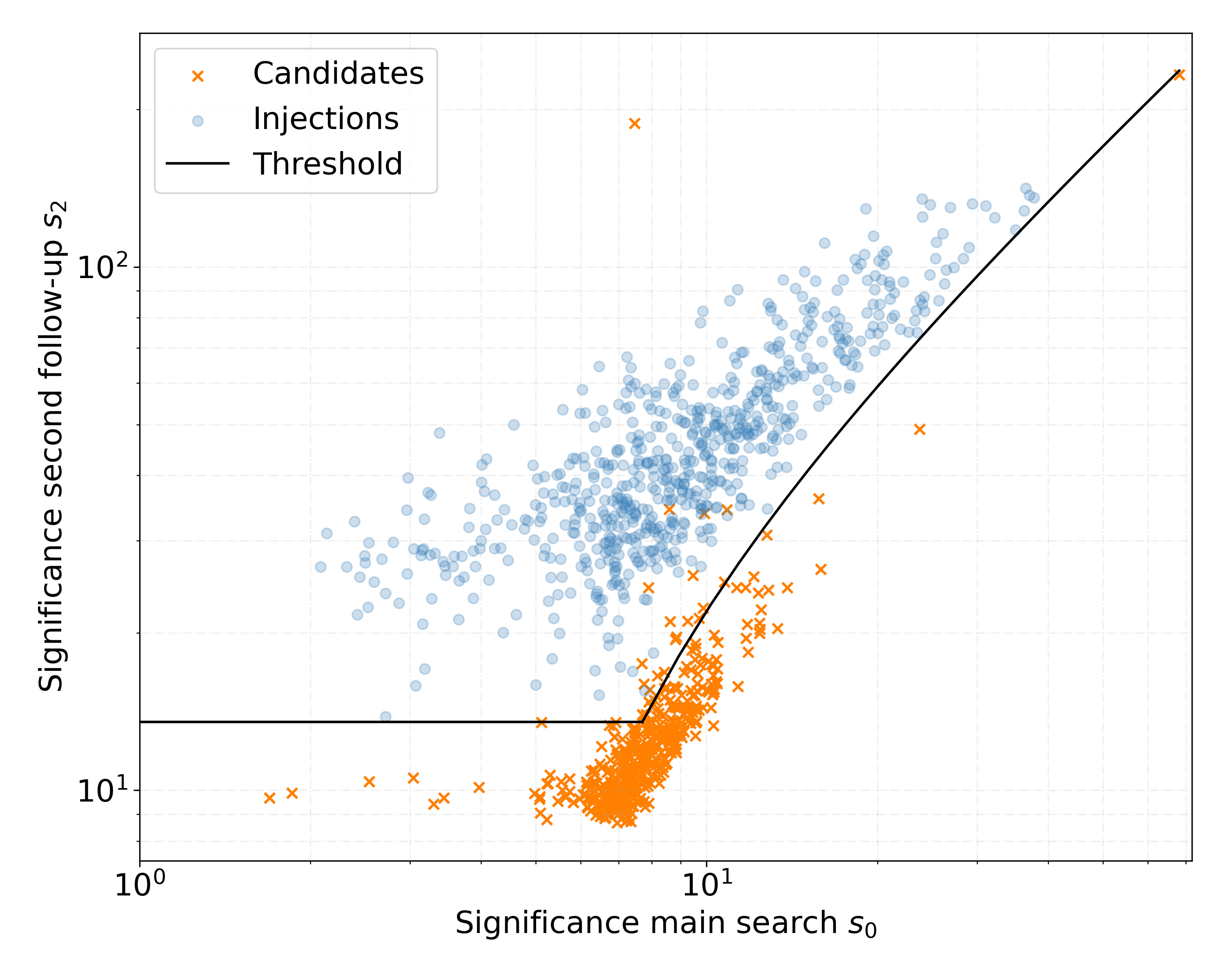}
    \caption{Orange crosses: significance of search candidates in stage 1 ($s_1$, left plot) and in stage 2 ($s_2$, right plot) as a function of their significance in stage 0 ($s_0$). Stage 1 investigates 18\,282 candidates. Stage 2 investigates 458 candidates. Blue crosses: target signal population with 1\,288 signals investigated by stage 1 of which 1\,284 survive and are investigated in stage 2. The candidate survival region is the region above the black line. We note that the loudest $s_2$ candidate out of stage 2 is due to a hardware injection of a signal from an isolated source, so a fake signal outside of the parameter space probed here. Consistently with this, the candidate is just below the acceptance threshold, and it is rejected. Its high $s_2$ value is due to the fake signal being very loud.}
    \label{fig:FU}
  \end{center}
\end{figure*}

If a candidate is due to a gravitational-wave signal we expect that its significance will increase through the different stages. Candidates from stage $i$ are discarded if their significance $s_i$ does not increase with respect to $s_0$ as expected for astrophysical signals. 
We calibrate this expectation with our target signal population, as shown in Figure~\ref{fig:FU}. 

\begin{figure}
  \begin{center}
    \includegraphics[width=1.0\columnwidth]{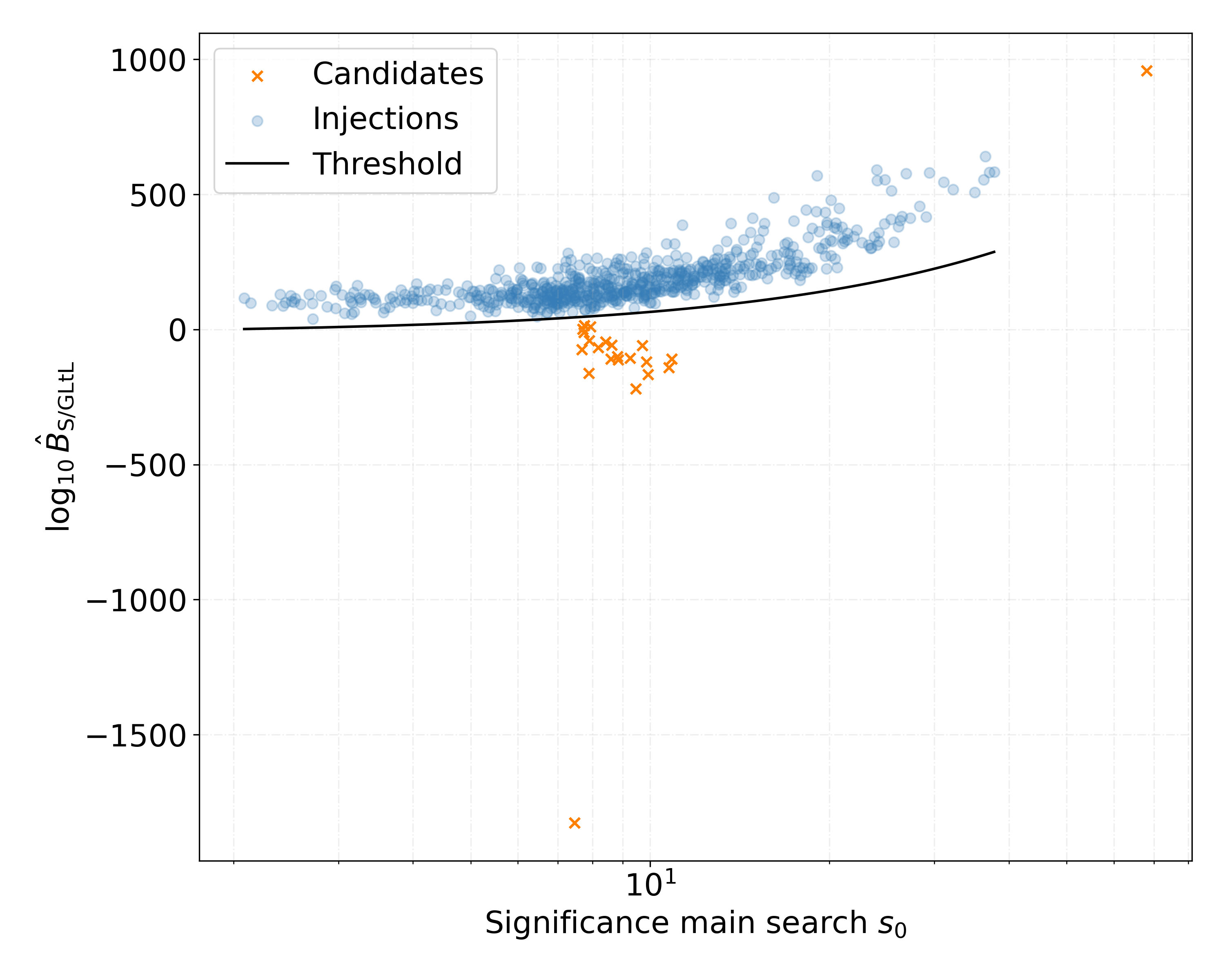}
    \caption{$\log_{10} \hat{B}_\mathrm{S/GLtL}$ values of the stage 2 candidates as a function of the stage 0 significance $s_0$. Blue circles: 1\,284 injected signals from the target signal population. Orange crosses: 24 surviving candidates from stage 2 plus the high-significance candidate at the upper right corner of the right plot in Figure~\ref{fig:FU}. The candidate survival region is the region above the black line. }
    \label{fig:BSGLtL2}
  \end{center}
\end{figure}

Only two follow-up stages are necessary to discard all candidates:
\begin{itemize}
\item {\bf{Stage 1}}: $T_{\mathrm{coh}} = 45\,000$\,s with a significance-veto false dismissal of $\simeq 0.3\%$ (see left plot of Figure~\ref{fig:FU}). This results in 458 surviving candidates out of 18\,282.

\item {\bf{Stage 2}}: $T_{\mathrm{coh}} = 135\,000$\,s with a significance-veto false dismissal of $< 0.1\%$ (see right plot of Figure~\ref{fig:FU}). This results in $24$ surviving candidates out of 458.

\item {\bf{$ \hat{B}_\mathrm{S/GLtL}$ veto}} with a false dismissal of $< 0.1\%$ (see Figure~\ref{fig:BSGLtL2}). No candidate survives out of the $24$ (the candidate above the black line is due to a hardware injection -- signals added to the data in hardware for pipeline validation purposes \citep{PhysRevD.95.062002}).
\end{itemize}

To investigate the nature of the $24$ surviving candidates we compute the line-robust statistic $\log_{10} \hat{B}_\mathrm{S/GLtL}$ with $\hat{\mathcal{F}}_* = 0$ as previously done, and compare it with what is expected from the target signal population as a function of $s_0$. The result is shown in Figure~\ref{fig:BSGLtL2}, and all candidates are vetoed.

Based on these results we conclude that none of our candidates can be associated with a real continuous wave signal.

\section{Results}
\label{sec:results}

\subsection{Estimated upper limits}
\label{sec:UL}

We estimate the population-averaged 95\%-confidence upper limits on the gravitational-wave amplitude $h_0^{95\%}(f)$ in every 0.1\,Hz band. This is the amplitude such that 95\% of a population of signals with that amplitude, frequency in that band, and with the same distribution as our target population, would have survived our pipeline. This means that these signals generate a cluster with a higher significance than the top cluster in the respective 5 mHz band, and the candidate passes the two stages of the follow-up and the $\hat{B}_\mathrm{S/GLtL}$ vetoes.

\begin{figure}
  \begin{center}
    \includegraphics[width=1.0\columnwidth]{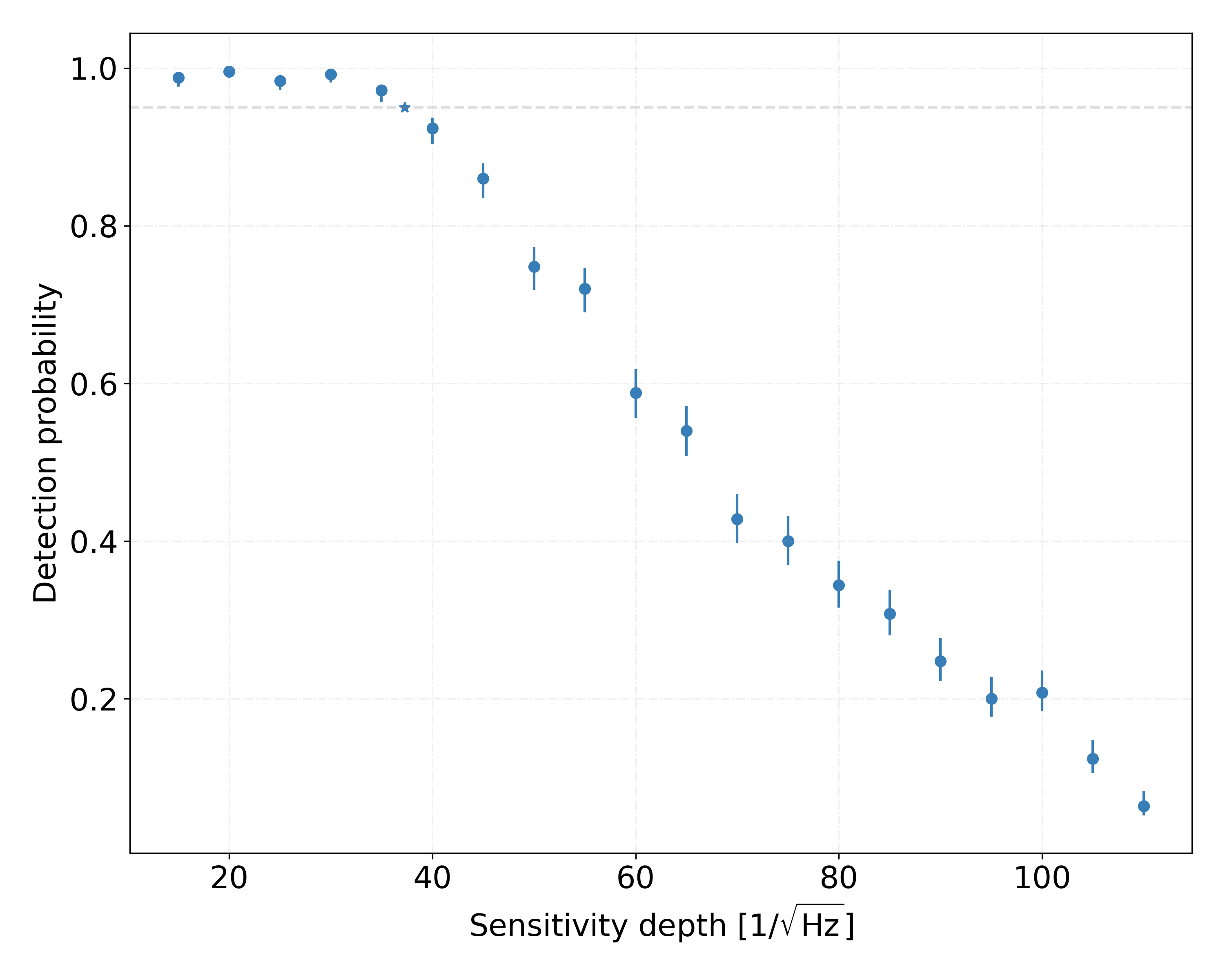}
    \caption{Detection probability as a function of the sensitivity depth $\sqrt{S_n(f_0)}/h_0$ of different signals with frequency $f_0$ (distributed as $f_0\in 85.55\pm 0.05$ Hz) and amplitude $h_0$. The error bars show the $1\sigma$ uncertainty, and the star marks the interpolated detection probability at $\mathcal{D}^{95\%}$.}
    \label{fig:sensdepth}
  \end{center}
\end{figure}

We determine the upper limits in the ten 0.1\,Hz bands mentioned in Section~\ref{sec:fakeSignals}. For each band and $h_0$ value we compute the fraction of detected signals out of a total of $250$ from our target population. We determine the $h_0$ corresponding to 95\% detection efficiency with a linear interpolation, as shown in Figure~\ref{fig:sensdepth}. 

We associate to every upper limit a sensitivity depth ${\mathcal{D}}$ \citep{Behnke:2014tma,Dreissigacker:2018afk}, defined as
\begin{equation}
\label{eq:sensDeph}
{\mathcal{D}}^{95\%}(f)=\frac{\sqrt{S_{\mathrm{n}}(f)}}{h_0^{95\%}(f)},
\end{equation}
where $\sqrt{S_{\mathrm{n}}(f)}$ is the amplitude spectral density of the data, shown in Figure~\ref{fig:PSD} with the middle green curve. Across the ten different bands the average 95\% sensitivity depth is $38.6 \pm 0.9$ $\textrm{Hz}^{-1/2}$, where the second value is the measured standard deviation. We use this value inverting Equation~\eqref{eq:sensDeph} to estimate the upper limits $h_0^{95\%}$ in the other frequency bands. This will yield correct upper limit estimates as long as the data is not affected by large disturbances.

We identify the 5 mHz sub-bands where the noise may be affected by large disturbances by looking at the significance of the loudest candidate. If it is too far (12 standard deviations) on the tails of the distribution in quiet bands, the upper limit does not hold in that sub-band \citep{O2aallskybinary}. The black lines shown in Figure~\ref{fig:Toplists} bracket the interval of significance values that define the accepted bands.

\begin{figure}
  \begin{center}
    \includegraphics[width=1.0\columnwidth]{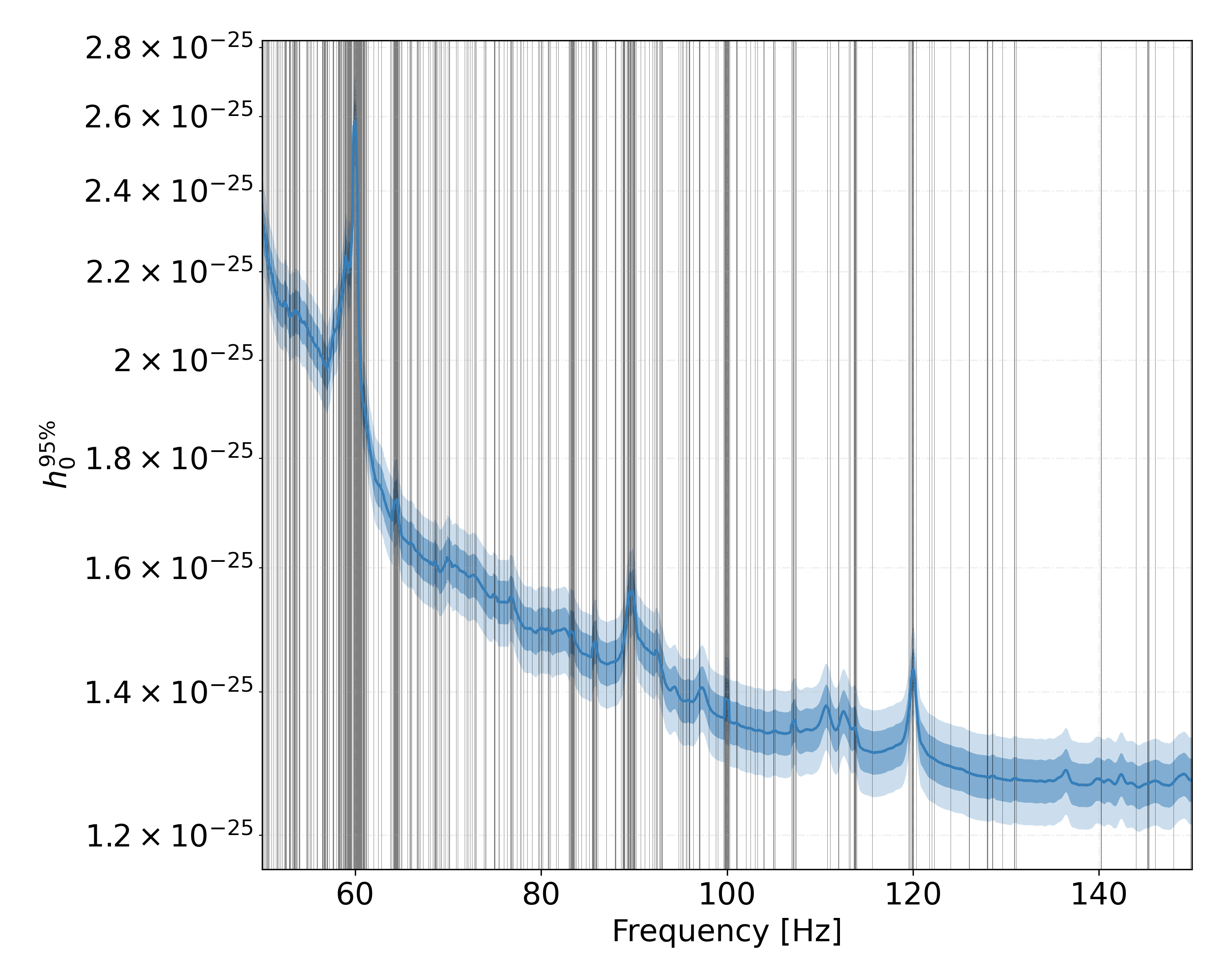}
    \caption{Estimated population-averaged upper limits on the gravitational-wave amplitude $h_0^{95\%}$ at the $95\%$ confidence level as a function of frequency. The shaded areas surrounding the curve indicate regions obtained with one and two measured standard deviations. The vertical black rectangles show the $3\,402$ (out of $20\,000$) 5 mHz frequency bands where the upper limits do not apply.}
    \label{fig:UL}
  \end{center}
\end{figure}

The estimated upper limit values and the list of frequency bands where the upper limits do not apply are available in machine-readable format in the supplementary materials\footnote{Upper limit values in machine-readable format and the list of vetoed frequency bands are available at \url{https://www.aei.mpg.de/continuouswaves/O3AllSkyBinary1} as well as from the journal.} and shown in Figure~\ref{fig:UL}. The most stringent upper limit is $1.25 \times 10^{-25}$ at $144.32$ Hz. Calibration uncertainties in the of the Advanced LIGO detectors affect this at a $\lesssim 10\%$ level over all frequencies \citep{Sun_2020, sun2021characterizationsystematicerroradvanced}.

The search of \cite{O3aallskybinaryLIGO} achieves a sensitivity depth of $\sim 22.5$ $\textrm{Hz}^{-1/2}$ at similar frequencies and in a different region of the binary parameter space, which is a factor $\sim 1.75$ times less sensitive than our result. We estimate that part of this additional sensitivity is due to our search using the entire O3 dataset, rather than just the first half as in \citet{O3aallskybinaryLIGO}. This accounts for a factor $\sim 2^{1/4}=1.2$. The parameter space covered in this search is about five times larger, quantified in terms of the breadth $\mathcal{B}$ given by Equation~(75) of \cite{WETTE2023102880}. Therefore, we conclude that this search method is intrinsically more sensitive by at least $\sim 46\%$.

If the continuous waves are sourced by an equatorial elliptiticty $\epsilon=|I_{xx} - I_{yy}|/I_{zz}$, with $I_{zz}$ being the moment of inertia of the star with respect to the principal axis aligned with the rotational axis, the gravitational-wave amplitude at a distance $d$ is 
\begin{equation}
 \label{eq:h0}
  h_0 = 10^{-26} \left[ \frac{ I_{zz} }{ 10^{38} \mathrm{~kg~m}^2} \right] \left[ \frac{\epsilon}{ 10^{-6} } \right]
                 \left[ \frac{f_0}{100 \mathrm{~Hz}} \right]^{2} \left[ \frac{1 \mathrm{~kpc}}{d} \right].
\end{equation}

\begin{figure}
  \begin{center}
    \includegraphics[width=1.0\columnwidth]{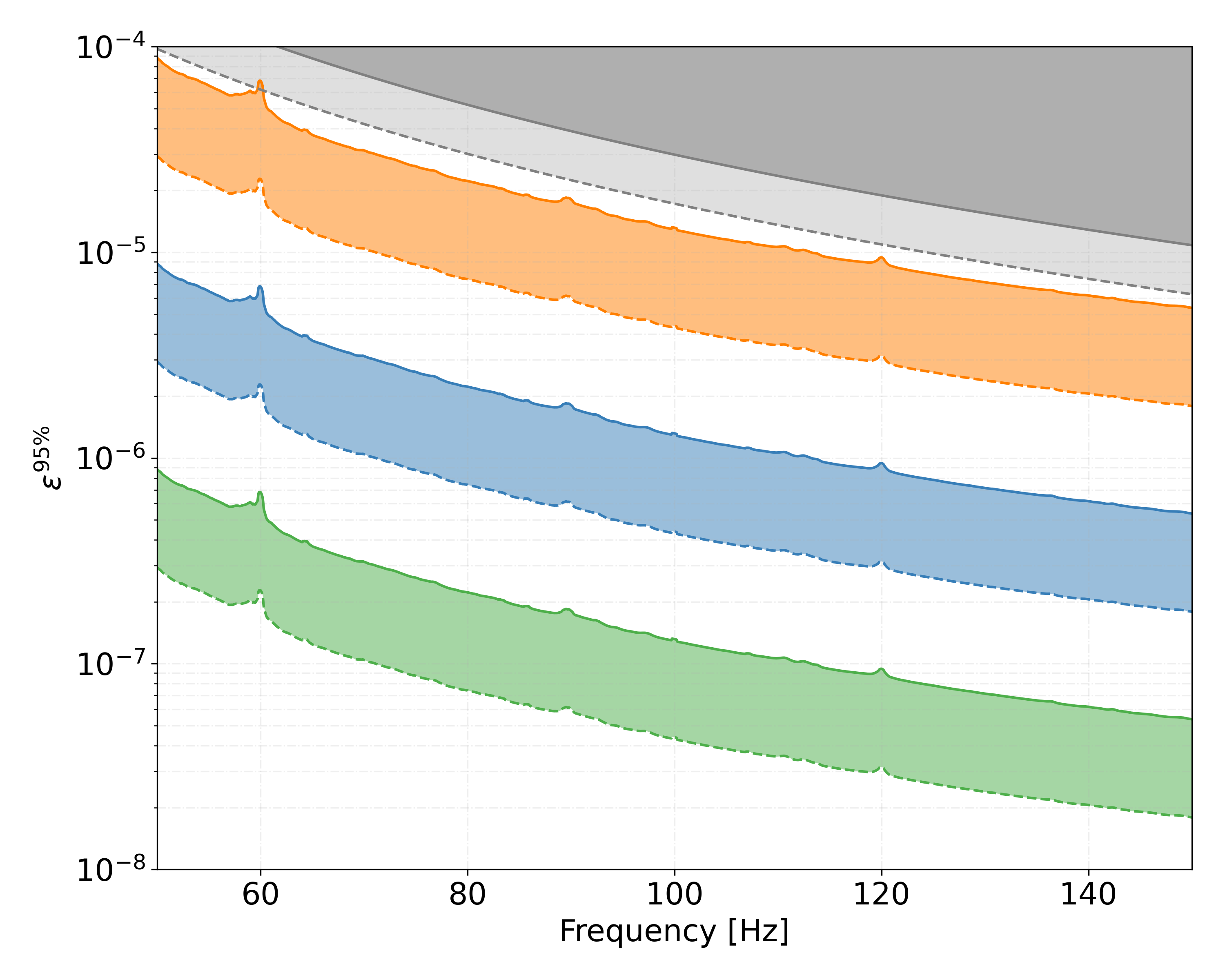}
    \caption{Estimated population-averaged upper limits on the neutron star ellipticity $\epsilon$ at the $95\%$ confidence level as a function of frequency. The three different colors show results for distances of 1\,kpc (upper orange curves), 100\,pc (middle blue curves), and 10\,pc (green bottom curves). The dashed lines use a moment of inertia of $I_{zz} = 3 \times 10^{38}$\,kg\,m$^2$ instead of the canonical $I_{zz} = 10^{38}$\,kg\,m$^2$ value, used by the solid lines. The measured standard deviation on the sensitivity depth propagated to the estimated upper limits is not large enough to be visible in the plot. The upper shaded gray areas show the regions where this search is not sensitive because the high ellipticities would generate a spin-down larger than the one probed by this search (light gray shows the limit for $I_{zz} = 3 \times 10^{38}$\,kg\,m$^2$, while dark gray shows the limit for $I_{zz} = 10^{38}$\,kg\,m$^2$).}
    \label{fig:Reach}
  \end{center}
\end{figure}

The estimated upper limits on $h_0$ can be used to estimate upper limits on the asymmetry of the targeted neutron star population by rearranging Equation~\ref{eq:h0}. These ellipticity upper limits depend on the moment of inertia of the neutron star, which is uncertain by around a factor of three \citep[see Section~4B of][]{known_2007}, although more exotic neutron star models such as quark stars or lower-mass neutron stars could have even higher moments of inertia \citep{Horowitz_2010,Owen_2005}. Figure~\ref{fig:Reach} shows these results at three different distances and two values of the moment of inertia. 

These upper limits on the ellipticity are the most restrictive to date for unknown neutron stars in binary systems. For sources at 1\,kpc, with $I_{zz} = 10^{38}$ kg m$^2$ and emitting continuous waves at 150\,Hz, the ellipticity can be constrained to be $ < 5.2 \times 10^{-6}$, while at 10 pc we have $\epsilon < 5.2 \times 10^{-8}$. If instead we assume $I_{zz} = 3 \times 10^{38}$ kg m$^2$ (as could be due to an equation of state that supports neutron stars with $\approx 70\%$ larger radii), the upper limits are more stringent, as shown by the dashed traces. For example, at 100\,pc and 100\,Hz, the upper limit is $\epsilon < 4 \times 10^{-7}$. If the true minimum ellipticity is $10^{-9}$ as suggested by \citet{Woan_Pitkin_Haskell_Jones_Lasky_2018}, we are about one order of magnitude above that limit for neutron stars at 10\,pc and 150 Hz, and two orders of magnitude for stars at 100\,pc.

If the continuous waves are sourced by r-modes, the gravitational-wave amplitude is given by \citep{Owen_2010}
\begin{equation}
  \begin{aligned}
    h_0 = & 10^{-26} \left[ \frac{\alpha}{2.8 \times 10^{-4}} \right] \left[ \frac{f_0}{100 \mathrm{~Hz}} \right]^{3} \left[ \frac{1 \mathrm{~kpc}}{d} \right] \\
          & \times  \left[ \frac{M}{1.4 ~{\mathrm{M}}_\odot} \right] \left[ \frac{R}{11.7 ~{\mathrm{km}}} \right]^3,
  \end{aligned}
  \label{eq:alpha}
\end{equation}
where $\alpha$ is the r-mode amplitude, $M$ is the mass of the neutron star, and $R$ its radius.

\begin{figure}
  \begin{center}
    \includegraphics[width=1.0\columnwidth]{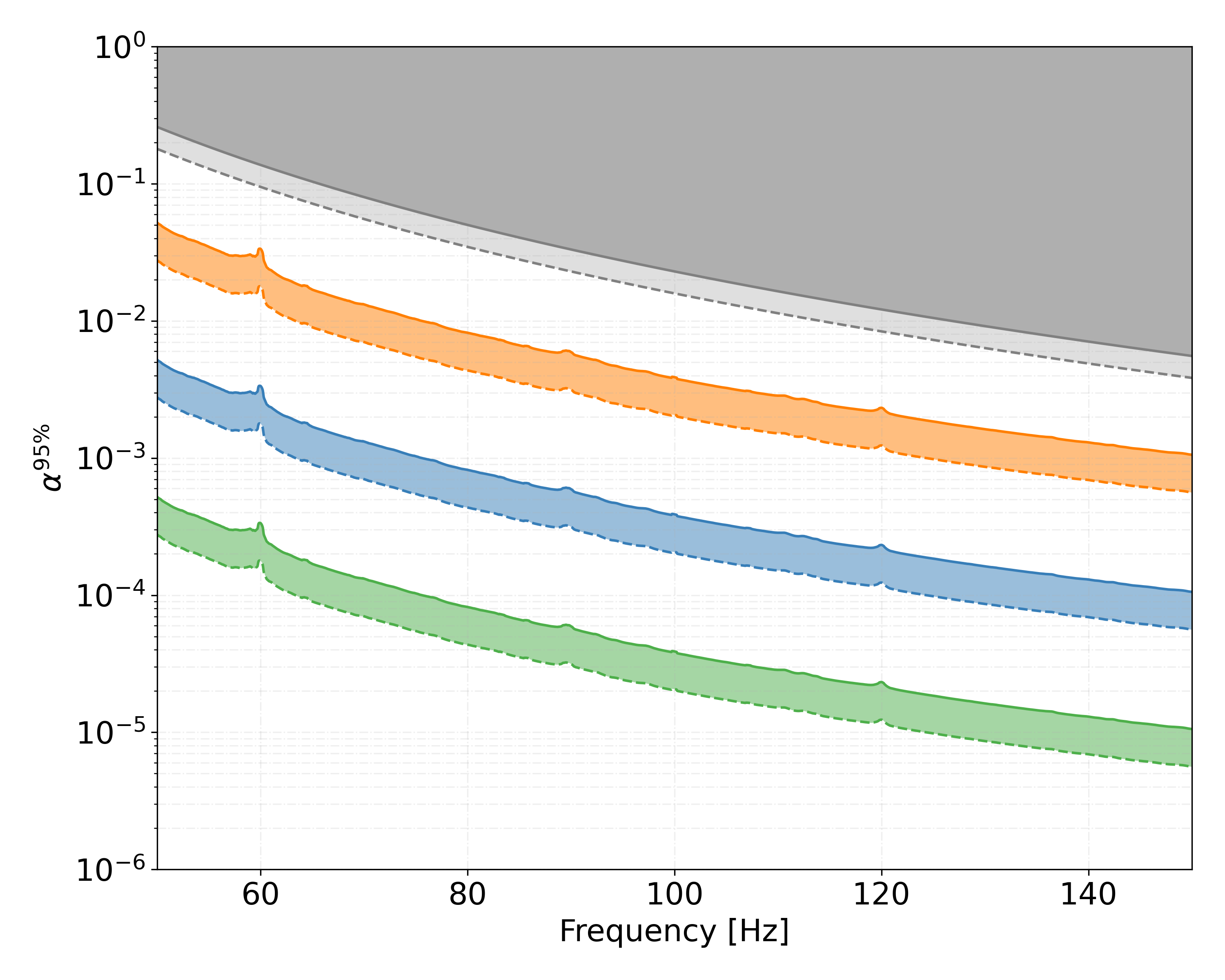}
    \caption{Estimated population-averaged upper limits on the r-mode amplitude $\alpha$ at the $95\%$ confidence level as a function of frequency. The three different colors show results for distances of 1\,kpc (upper orange curves), 100\,pc (middle blue curves), and 10\,pc (green bottom curves). The dashed lines assume $M=1.9 ~M_\odot$ and $R=13$ km, while the solid lines assume $M=1.4 ~M_\odot$ and $R=11.7$ km. The measured standard deviation on the sensitivity depth propagated to the estimated upper limits is not large enough to be visible in the plot. The upper shaded gray areas show the regions where this search is not sensitive because the high r-mode amplitudes would generate a spin-down larger than the one probed by this search (light gray shows the limit for $M=1.9 ~M_\odot$ and $R=13$ km, while dark gray shows the limit for $M=1.4 ~M_\odot$ and $R=11.7$ km).}
    \label{fig:ReachRModes}
  \end{center}
\end{figure}

The estimated upper limits on $h_0$ can be used to estimate upper limits on the r-mode amplitude of the targeted neutron star population by rearranging Equation~\ref{eq:alpha}. These r-mode amplitude upper limits depend on the neutron star mass and radius (connected by the unknown equation of state). Figure~\ref{fig:ReachRModes} shows these results at three different distances and two mass-radius combinations.

For sources at 1\,kpc, with $M=1.4 ~M_\odot$, $R=11.7$ km and emitting continuous waves at 150\,Hz, the r-mode amplitude can be constrained at $\alpha < 10^{-3}$, while at 10\,pc we have $\alpha < 10^{-5}$. If instead we assume $M=1.4 ~M_\odot$ and $R=11.7$ km, the upper limits are more stringent, as shown by the dashed traces. For example, at 100\,pc and 100\,Hz, the upper limit is $\alpha < 2 \times 10^{-4}$. The range of theoretical predictions in the literature on r-mode amplitudes is $\alpha\sim 8\times10^{-7}-10^{-4}$ \citep{Gusakov:2013jwa,gusakov_explaining_2014} or $\alpha\sim10^{-5}$ \citep{bondarescu_spin_2007}. It can be seen that for neutron stars between 10 and 100 pc, our results start to constrain the upper limits of these theoretical ranges.

\subsection{Robustness of results}
\label{sec:robustness}

The signal model described in Section \ref{sec:signal} and commonly used in CW searches assumes phase coherence of the signal for the duration of each coherent segment. While based on the observations of millisecond pulsars this is a very reasonable assumption for fast spinning neutron stars, it is known that some pulsars exhibit glitches \citep{PhysRevD.96.063004}, spin-wandering \citep{spinwandering}, and timing noise \citep{PhysRevD.91.062009}. The phenomenology of these effects is rich and diverse, and the underlying causes not completely clear.

A loss of phase coherence will impact the sensitivity of a search if it is resolvable by it. For a coherent search over a time $T_{\mathrm{coh}}$ the maximum frequency shift $\Delta f$ that does not produce a significant loss in detection efficiency is 
\begin{equation}
\Delta f \lesssim \frac{1}{ T_{\mathrm{coh}} }.
\label{eq:maxDeltaF}
\end{equation}
For this search the largest coherence time is used in the last stage of the follow-up, and it is in that stage that signals with deviations from the perfectly phase-coherent model are more likely to be less effectively detected compared to the the standard model signals. The last follow-up stage has a coherence time $T_{\mathrm{coh}} = 1.35\times 10^5$ s, which implies $\Delta f \lesssim 7.4~\mu$Hz.

Based on assessments of the varying accretion torque from X-ray observations of Sco X-1, \cite{spinwandering} have estimated the likelihood of the magnitude of the wandering in frequency $\Delta f$ for observations of various durations. Figure 7 of \cite{spinwandering} applies to the frequency range of interest for this search and clearly shows that variations $\geq 7.4~\mu$Hz are extremely unlikely for $T_{\mathrm{coh}} = 1.35\times 10^5$ s and even over a year more than 97\% of the simulated population present spin-wandering smaller than $7.4~\mu$Hz. Although this search does not target emission from Sco X-1, since Sco X-1 is the source with the highest known accretion rate it represents a ``worst case scenario" for spin-wandering: for the sources we target here we expect a lower accretion rate and a smaller frequency variation than for Sco X-1.

Regarding timing noise, figure 5 of \cite{PhysRevD.91.062009} shows that for $T_{\mathrm{coh}} \approx$ a day, the mismatch is much smaller than $1\%$, so we expect that the impact of timing noise on our longest $T_{\mathrm{coh}}$ -- which is around this order of magnitude -- to be negligible (calibration amplitude uncertainties are an order of magnitude larger).

The likelihood of occurrence of glitches in pulsars has been studied by \cite{PhysRevD.96.063004}. For objects with low spin-down, such as the ones that we target here, the top panel of figure 8 in \cite{PhysRevD.96.063004} shows that the probability of observing a glitch on the time scale of a day is $\lesssim 1\%$. It grows to about $25\%$ over the total observation time and having factored-in the duty factor. This means that a glitch is $\approx 25\%$ likely occur in one of the segments, assuming the highest spin-down allowed by our search. However, figure 4 in \cite{PhysRevD.96.063004} shows that Crab-like glitches are expected to be small enough ($< 7.4~\mu$Hz) to produce a negligible mismatch (this is also indicated in their figure 6).

Our search targets signals with waveforms described in Section \ref{sec:signal}. Signals that deviate from our model can always be conceived, for which the detection efficiency of this search might be lower compared to the model described in Section \ref{sec:signal}. To address this possibility, other searches exist that make fewer assumptions on the signal waveform. Such searches are more robust but also less sensitive to the standard signals, see for instance \cite{PhysRevD.93.123009,PhysRevD.96.102006,PhysRevD.97.043013,PhysRevD.104.042003,PhysRevD.100.023006}. For this search we chose a different approach. However, timing noise, glitches, and spin-wandering are {\it{observed}} phenomena in neutron stars, and their impact on CW searches has been studied. Based on those studies we argue that our search is reasonably robust to deviations of the signal waveform due to those effects.

Finally, we have only tested the efficiency of our search with signals distributed as explained in section \ref{sec:fakeSignals}. If a putative signal lies outside our parameter-space ranges, then the false dismissal probability that we have estimated would be increased. Quantifying this is outside of the scope of this paper.

\section{Conclusions}
\label{sec:conclusions}

%Summary
In this paper we have presented results from a direct search for continuous gravitational waves from neutron stars in binary systems with the largest orbital periods and widest orbits ever directly investigated. We set the most constraining direct upper limits in the investigated region of parameter space.

This search exemplifies the capabilities of the newly developed \textsc{BinarySkyHou$\mathcal{F}$} pipeline, thanks to the usage of a longer coherence time, a more sensitive detection statistic, and finer parameter space grids in both the coherent and semi-coherent grids. The search took $\sim 82\,000$ CPU-hours to complete, running on a combination of Intel(R) Xeon(R) Platinum 8360Y @ 2.40GHz CPUs and NVIDIA A100-SXM4-40GB GPUs. We estimate the raw sensitivity improvement from this pipeline to be at least $46\%$ compared to the most sensitive previous results.

\cite{Singh:2019han} proposed that searches for continuous waves from isolated neutron stars also probe signals from neutron stars in high orbital period binary systems with small eccentricity. We refer the reader to that paper for the details. \cite{Singh:2022hfd} applied this idea to the null results of the search for continuous waves from isolated neutron stars by \cite{Steltner:2020hfd} and based on that constrained the gravitational wave amplitude with a high sensitivity depth of $\sim 50$ Hz$^{-1/2}$. Their $( P_\mathrm{orb}, a_{\mathrm{p}} )$ parameter range partly overlaps the parameter space that we have investigated here, as shown in Figure~\ref{fig:ParameterSpace}. Their investigation is however not a direct search and can only recast a null result to the binary parameter space, or re-interpret a detection ascribed to an isolated system as coming from a binary. Conversely, the search presented here is a direct search and could detect a new signal.

\begin{acknowledgments}

This project has received funding from the European Union’s Horizon 2020 research and innovation program under the Marie Sklodowska-Curie Grant Agreement No. 101029058.

This work has utilized the ATLAS cluster computing at MPI for Gravitational Physics Hannover and the HPC system Raven at the Max Planck Computing and Data Facility.

This research has made use of data or software obtained from the Gravitational Wave Open Science Center (\url{gwosc.org}), a service of the LIGO Scientific Collaboration, the Virgo Collaboration, and KAGRA \citep{GWOSC}. This material is based upon work supported by NSF's LIGO Laboratory which is a major facility fully funded by the National Science Foundation, as well as the Science and Technology Facilities Council (STFC) of the United Kingdom, the Max-Planck-Society (MPS), and the State of Niedersachsen/Germany for support of the construction of Advanced LIGO and construction and operation of the GEO600 detector. Additional support for Advanced LIGO was provided by the Australian Research Council. Virgo is funded, through the European Gravitational Observatory (EGO), by the French Centre National de Recherche Scientifique (CNRS), the Italian Istituto Nazionale di Fisica Nucleare (INFN) and the Dutch Nikhef, with contributions by institutions from Belgium, Germany, Greece, Hungary, Ireland, Japan, Monaco, Poland, Portugal, Spain. KAGRA is supported by Ministry of Education, Culture, Sports, Science and Technology (MEXT), Japan Society for the Promotion of Science (JSPS) in Japan; National Research Foundation (NRF) and Ministry of Science and ICT (MSIT) in Korea; Academia Sinica (AS) and National Science and Technology Council (NSTC) in Taiwan.

\end{acknowledgments}

\bibliography{Paper}{}
\bibliographystyle{aasjournal}

\end{document}